\documentclass[twocolumn, tighten, twocolappendix]{aastex631}

\usepackage{xcolor}

\begin{document}

\title{The Non-Eruptive Reconfiguration of a Quiescent Filament After a Nearby Active Region Emergence}

\correspondingauthor{James McKevitt}
\email{james.mckevitt.21@ucl.ac.uk}

\author[0000-0002-4071-5727]{James McKevitt}
\affiliation{University College London, Mullard Space Science Laboratory, Holmbury St Mary, Dorking, Surrey, RH5 6NT, UK}
\affiliation{University of Vienna, Institute of Astrophysics, Türkenschanzstrasse 17, Vienna A-1180, Austria}

\author[0000-0001-9457-6200]{Louise Harra}
\affiliation{Physikalisch-Meteorologisches Observatorium Davos/World Radiation Center, PMOD/WRC, Dorfstrasse 33, Davos Dorf, 7260, GR, Switzerland}
\affiliation{D-PHYS, ETH Zürich, Wolfgang-Pauli Strasse 27, Zürich, 8093, ZH, Switzerland}

\author[0000-0001-7809-0067]{Gherardo Valori}
\affiliation{Max-Planck-Institut für Sonnensystemforschung, Justus-vonLiebig-Weg 3, 37077 Göttingen, Germany}

\author[0000-0002-0665-2355]{Deborah Baker}
\affiliation{University College London, Mullard Space Science Laboratory, Holmbury St Mary, Dorking, Surrey, RH5 6NT, UK}

\author{Nils Janitzek}
\affiliation{D-PHYS, ETH Zürich, Wolfgang-Pauli Strasse 27, Zürich, 8093, ZH, Switzerland}
\affiliation{Physikalisch-Meteorologisches Observatorium Davos/World Radiation Center, PMOD/WRC, Dorfstrasse 33, Davos Dorf, 7260, GR, Switzerland}

\author[0000-0003-2802-4381]{Stephanie Yardley}
\affiliation{Department of Mathematics, Physics and Electrical Engineering, Northumbria University, Ellison Place, Newcastle upon Tyne, NE1 8ST, UK}
\affiliation{University College London, Mullard Space Science Laboratory, Holmbury St Mary, Dorking, Surrey, RH5 6NT, UK}
\affiliation{Donostia International Physics Center (DIPC), Paseo Manuel de Lardizabal 4, 20018 San Sebasti{\'a}n, Spain}

\author[0000-0001-9346-8179]{Sarah Matthews}
\affiliation{University College London, Mullard Space Science Laboratory, Holmbury St Mary, Dorking, Surrey, RH5 6NT, UK}

\author[0000-0002-6287-3494]{Hamish Reid}
\affiliation{University College London, Mullard Space Science Laboratory, Holmbury St Mary, Dorking, Surrey, RH5 6NT, UK}

\author[0000-0001-7927-9291]{Alexander W. James}
\affiliation{University College London, Mullard Space Science Laboratory, Holmbury St Mary, Dorking, Surrey, RH5 6NT, UK}

\author[0000-0002-8538-3455]{Muriel Stiefel}
\affiliation{University of Applied Sciences and Arts Northwestern Switzerland, Bahnhofstrasse 6, 5210 Windisch, Switzerland}
\affiliation{D-PHYS, ETH Zürich, Wolfgang-Pauli Strasse 27, Zürich, 8093, ZH, Switzerland}

\author[0000-0002-2189-9313]{David H. Brooks}
\affiliation{Computational Physics Inc., Springfield, VA 22151, USA}
\affiliation{University College London, Mullard Space Science Laboratory, Holmbury St Mary, Dorking, Surrey, RH5 6NT, UK}

\author[0000-0003-4437-0698]{Ryan Dewey}
\affiliation{Department of Climate and Space Sciences and Engineering, University of Michigan, 2455 Hayward Street, Ann Arbor, MI 48109-2143, USA}

\author[0000-0001-5956-9523]{Jim M. Raines}
\affiliation{Department of Climate and Space Sciences and Engineering, University of Michigan, 2455 Hayward Street, Ann Arbor, MI 48109-2143, USA}

\author[0000-0003-1611-227X]{Susan T. Lepri} 
\affiliation{Department of Climate and Space Sciences and Engineering, University of Michigan, 2455 Hayward Street, Ann Arbor, MI 48109-2143, USA}

\author[0000-0002-5975-7476]{Liang Zhao}
\affiliation{Department of Climate and Space Sciences and Engineering, University of Michigan, 2455 Hayward Street, Ann Arbor, MI 48109-2143, USA}

\author[0000-0003-4319-2009]{{Juan~Sebasti\'an} {Castellanos~Dur\'an}}
\affiliation{Max-Planck-Institut für Sonnensystemforschung, Justus-vonLiebig-Weg 3, 37077 Göttingen, Germany}

\begin{abstract}

The unpredictability of solar filament eruptions presents major challenges for forecasting space weather, as such eruptions frequently drive coronal mass ejections (CMEs) that impact the heliosphere. While nearby flux emergence is often linked to their destabilisation, the specific characteristics of both the emerging flux and the filament that determine whether an eruption occurs remain unclear. We report observations of a quiescent filament that did not erupt following the nearby emergence of active region NOAA~13270 and a subsequent C-class flare in April 2023. Our analysis combines multi-viewpoint extreme ultraviolet (EUV) imaging and X-ray imaging with EUV spectroscopy, radio imaging and measurements of, and extrapolations from, the photospheric magnetic field. We identify the formation of a coronal null point and fan-spine topology at the interface between the active region and filament which exhibited persistent slow reconnection, indicated by chromospheric brightenings, persistent radio emission, and plasma upflows. Our results indicate that ongoing reconnection and jets can relieve magnetic stress and enable filament stability, even when under strong perturbation. We suggest that the orientation of emerging flux relative to the ambient field is a critical parameter in filament evolution, and provide observational constraints for models of filament stability and eruption.

\end{abstract}

\section{Introduction}

Solar filaments, cool and dense structures of plasma in the hotter and more tenuous corona, are a key element of solar activity \citep{vial_solar_1998, parenti_solar_2014}. The filament plasma resides within filament channels, structures of highly sheared and/or twisted magnetic fields located above photospheric polarity-inversion lines \citep{martin_conditions_1998, wang_formation_2007}. The formation of these filament channels is strongly influenced by magnetic shear at the photospheric level \citep{schmieder_differential_1996}. Filaments can become destabilised and erupt, yet they may also withstand considerable photospheric forcing without producing coronal mass ejections (CMEs), making predictions of their eruptive behaviour particularly challenging.

Quiescent filaments, typically found far from active regions, are rooted in large-scale, weaker fields associated with decayed active regions \citep{engvold_observations_1998, vial_formation_2015}. These filaments experience gradual coronal field stress due to surface shear motions, episodic flux emergence, or flux cancellation, accumulating free magnetic energy that is eventually released either gradually via smaller-scale reconnection events or explosively through large-scale eruptions \citep{antiochos_model_1999}. A significant driver of filament activity is the emergence of new magnetic flux associated with active regions.

Observations frequently highlight the crucial role of flux emergence in destabilising existing coronal structures, including large quiescent filaments and their arcades \citep[e.g.,][]{feynman_initiation_1995}. Both statistical analyses and detailed case studies have demonstrated that the formation of a nearby active region often triggers filament eruptions through magnetic interactions between emerging flux and the ambient magnetic environment \citep[e.g.,][]{balasubramaniam_disappearing_2011}. Magnetic reconnection resulting from these interactions can effectively remove, restructure, or reorient overlying magnetic fields, allowing filament channels to rise and erupt \citep{feynman_initiation_1995, wang_filament_1999, chen_emerging_2000}.

The emergence of magnetic flux can also drastically alter the coronal magnetic field topology, creating complex magnetic structures such as fan-spine configurations \citep{torok_fanspine_2009, duan_formation_2024}. In their simpler configuration, fan-spine topologies are characterised by a dome-shaped separatrix surface (fan) anchored at the photosphere, with a null point --- where the magnetic field strength is zero --- located at the dome apex. Inner and outer spine field lines intersect this null point. The inner spine is typically rooted in a photospheric parasitic polarity, that being flux opposite in sign to its surrounding magnetic environment. Such coronal null points are stable structures which can last for days \citep{luoni_magnetic_2007}, and are preferential sites for reconnection \citep{pontin_generalised_2011}. 

Fan-spine reconnection is associated with a variety of solar phenomena, including jets, circular flare ribbons, remote flaring activity, and radio bursts \citep[e.g.,][]{masson_nature_2009, wang_circular_2012, reid_x-ray_2012, liu_circular-ribbon_2015, li_two_2018, duan_homologous_2022, duan_formation_2024}. Observationally, persistent upflows of coronal plasma often mark sites of magnetic reconnection occurring at separatrix footprints and quasi-separatrix layers (QSLs). \cite{edwards_comparison_2016} demonstrated that such upflows frequently coincide spatially with photospheric footprints of null-point separatrices. Complementary work by \cite{demoulin_3d_2013} suggests that these upflows arise from a pressure gradient established between reconnecting high-pressure active-region loops and adjacent lower-pressure loops along QSLs. Additional numerical studies indicate similar plasma outflows can be driven simply by active-region expansion \citep{murray_outflows_2010}.

Several mechanisms are proposed for magnetic interactions triggering filament eruptions \cite[see Table~1 of][]{green_origin_2018}. In the magnetic breakout model, reconnection occurs at a magnetic null point between a sheared filament-channel arcade and adjacent magnetic fields in a multipolar (typically quadrupolar) configuration, removing the overlying, unsheared (strapping) fields and enabling eruption of the core structure \citep{antiochos_magnetic_1998, antiochos_model_1999, devore_homologous_2008, lynch_topological_2008}. Alternatively, eruption can be triggered by internal tether-cutting reconnection beneath the filament within the sheared core field, which is often facilitated by flux cancellation at the polarity inversion line \citep{moore_onset_2001, sterling_evidence_2010, aulanier_formation_2010}. In some cases, suitably oriented magnetic flux emergence near the polarity inversion line may promote either internal or external reconnection, injecting axial flux or currents and destabilising the filament channel \citep{feynman_initiation_1995, chen_emerging_2000, li_filament_2015, dacie_sequential_2018, torok_solar_2024}. Previous studies have found large-scale filament eruptions associated with nearby flux emergence and cancellation but the precise mechanism, whether breakout, internal tether-cutting, or ideal magneto-hydrodynamic instability, may vary or act in combination in individual events \citep{sterling_evidence_2010, aulanier_formation_2010}.

Observational studies typically correlate large-scale filament eruptions with nearby flux emergence events, particularly when mutual field orientations favour reconnection \citep{feynman_initiation_1995, wang_filament_1999, torok_solar_2024}. Nevertheless, several counterexamples highlight that filaments can endure significant magnetic perturbations without erupting \citep[e.g.,][and references therein]{li_filament_2015, torok_solar_2024}.

Consequently, non-eruptive interactions between emerging active regions and quiescent filament channels serve as crucial observational tests for existing eruption models. Understanding why some stressed magnetic configurations remain stable, while others lead to eruptions, CMEs, or both, requires detailed case studies connecting photospheric magnetic evolution, coronal magnetic field topology, and observational signatures of magnetic reconnection. 

In this study, we examine one such non-eruptive example between 1~April and 7~April~2023, whereby a bipolar active region (NOAA~13270) emerged adjacent to a pre-existing filament channel, creating a null point and associated separatrix dome between the newly formed and surrounding ambient fields. We investigate the corresponding signatures of reconnection, including persistent plasma upflows, electron beam indicators, and notable filament reconfiguration. We also analyse the largest flare from this active region --- a C-class flare on 6~April --- which is not associated with the eruption of the filament.

The paper is structured as follows. In Section~\ref{sec:observations} we present an overview of the observations and instruments we use in this study. In Section~\ref{sec:results} we show in detail the emergence of the active region nearby the filament and the resulting activity and magnetic topology. In Section~\ref{sec:discussion} we discuss the changes in magnetic topology during the interaction and explore the reasons the filament was not destabilised. In Section~\ref{sec:conclusions} we present our conclusions.

\section{Overview of observations}\label{sec:observations}

\begin{figure}
  \centering
  \includegraphics[width=.9\linewidth]{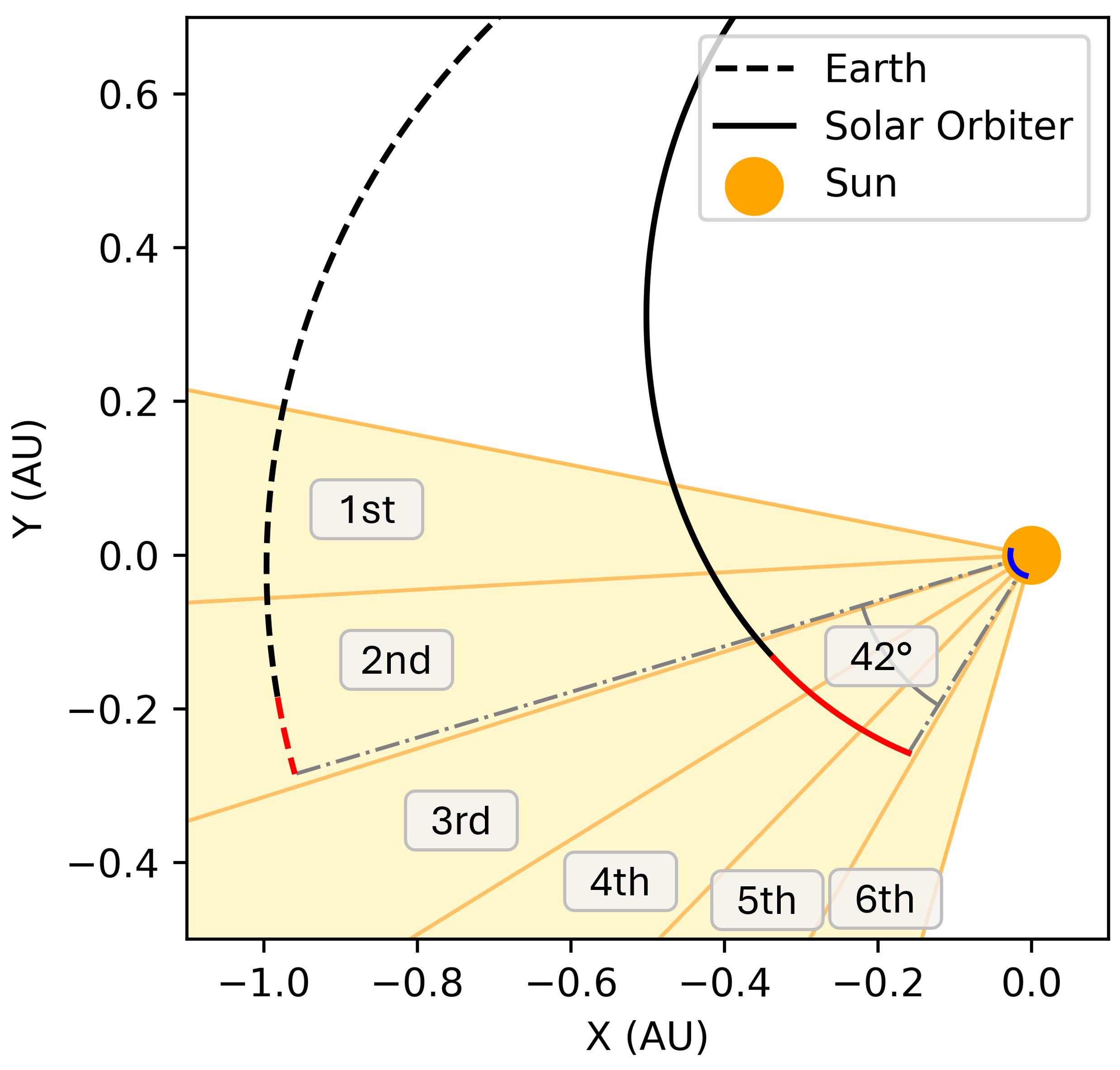}
  \caption{The positions of Solar Orbiter and Earth, viewed from above the ecliptic plane. Their locations for the period from 00:00~UT on 1~April to 00:00~UT on 7~April~2023 are highlighted in red. The active region's path on the solar surface between these times is shown in blue, and a yellow line projects radially from the centre of the active region for the dates of the study.}
  \label{fig:sc_locs}
\end{figure}

\begin{table*}
  \centering
  \caption{Summary of instruments used in this study.}
  \label{tab:instruments}
  \begin{tabular}{llllll}
    \hline
     & Instrument & Platform & Location / Orbit & Function & Observable \\
    \hline
    \multicolumn{6}{l}{\textbf{Ground-based}} \\
    & Radioheliograph & Nançay Radio Obs. & Earth & Radio imaging & Particle acceleration \\
    \hline
    \multicolumn{6}{l}{\textbf{Earth-orbiting}} \\
    & AIA & SDO & Geosynchronous & EUV imaging & Atmospheric structure and evolution \\
    & HMI & SDO & Geosynchronous & Magnetograms & Photospheric magnetic field \\
    & XRS & GOES-16 & Geostationary & X-ray flux & Flare timing \\
    & EIS & Hinode & Sun-synchronous & EUV spectroscopy & Coronal plasma flows \\
    & XRT & Hinode & Sun-synchronous & Soft X-ray imaging & High energy release \\
    \hline
    \multicolumn{6}{l}{\textbf{Sun-orbiting}} \\
    & EUI-FSI & Solar Orbiter & $\leq$42\textdegree{} ahead of Earth & EUV imaging & Atmospheric structure and activity \\
    & STIX & Solar Orbiter & $\leq$42\textdegree{} ahead of Earth & Hard X-ray imaging & High energy release \\
    \hline
  \end{tabular}
\end{table*}

We used ground-based, Earth-orbiting, and Sun-orbiting instruments in our study. These are summarised in Table~\ref{tab:instruments} and detailed in this section.

\subsection{EUV Imaging: Coronal structure and evolution}

We used multi-wavelength EUV imaging to track the temporal evolution of coronal structures and identify signatures of magnetic reconnection. The Atmospheric Imaging Assembly onboard the Solar Dynamics Observatory \citep[SDO/AIA;][]{pesnell_solar_2012,lemen_atmospheric_2012} provided continuous full-disk observations from Earth's perspective for the duration of our study. SDO/AIA's multiple temperature channels enabled us to distinguish between cool filament material (304~\AA{}; \(\log{T\sim{}4.7}\)), coronal loops at various temperatures (171~\AA{}; \(\log{T\sim{}}\)5.8, 193~\AA{}; \(\log{T\sim{}}\)6.2 and 7.3), and hot flare plasma (94~\AA{}; \(\log{T\sim{}6.8}\)).

Complementary EUV imaging from the Extreme Ultraviolet Full-Sun Imager onboard Solar Orbiter \citep[SO/EUI-FSI;][]{muller_solar_2020,rochus_solar_2020} provided a second viewpoint up to 42~degrees ahead of Earth relative to the Sun's rotation (shown in Figure~\ref{fig:sc_locs}). This enabled us to resolve foreshortening effects that affected Earth-based observations later in the study, particularly for determining the spatial relationship between the active region and filament structures. We used the 174~\AA{} and 304~\AA{} channels, which observe mostly $\sim$1~MK coronal plasma and $\sim$0.1~MK chromospheric and transition region plasma respectively \citep[e.g.,][]{chen_transient_2021,hayes_high-energy_2024}.

\subsection{EUV Spectroscopy: Plasma flows}

To identify and quantify plasma flows indicative of magnetic reconnection, we analysed spectroscopic observations from the EUV Imaging Spectrometer on Hinode \citep[Hinode/EIS;][]{kosugi_hinode_2007,culhane_euv_2007}. We focused primarily on the strong \ion{Fe}{12}~195.119~\AA{} ($\sim$1.6~MK) coronal emission line, which provides reliable Doppler measurements in typical active region plasma. We also analysed observations at higher temperatures using the \ion{Fe}{16}~262.984~\AA{} emission line ($\sim$2.5~MK).

We derived line-of-sight Doppler velocities by fitting Gaussian profiles to the emission lines using the \texttt{MPFIT} algorithm \citep{markwardt_non-linear_2009} implemented in the EIS Python Analysis Code \citep[EISPAC;][]{weberg_eispac_2023}. The Doppler velocity \(v\) was calculated from the fitted line centroid \(\lambda\), rest wavelength \(\lambda_0\), and speed of light \(c\) using \(v=c\frac{\lambda-\lambda_0}{\lambda_0}\).

\subsection{Photospheric Magnetograms: Magnetic field observations and extrapolations}

We tracked the emergence and evolution of photospheric magnetic flux using magnetograms from the Helioseismic and Magnetic Imager on SDO \citep[SDO/HMI;][]{hoeksema_helioseismic_2014}. Where vector magnetic field data were used, these were disambiguated using the minimum energy method \citep{metcalf_resolving_1994,metcalf_overview_2006} for active region fields and the radial acute method for quiet Sun regions.

To understand the three-dimensional coronal magnetic topology and identify sites of potential reconnection, we performed non-linear force-free field (NLFFF) extrapolations using the photospheric vector magnetograms. Details of our vector potential-based NLFFF method are provided in the Appendix. We identified quasi-separatrix layers (QSLs) --- volumetric structures where magnetic connectivity changes rapidly --- by calculating squashing factor $Q$-maps from our extrapolated fields using the \texttt{QSL Squasher} implementation \citep{tassev_qsl_2017}. High $Q$ values indicated strong field line divergence characteristic of separatrix surfaces and potential reconnection sites.

\subsection{Radio Imaging: Particle acceleration}

To identify sites of persistent particle acceleration associated with magnetic reconnection, we analysed radio imaging from the Nançay Radioheliograph \citep[NRH;][]{trottet_nancay_1997}. The NRH's interferometric observations at frequencies between 150--445~MHz enabled us to track type I noise storms, which indicate quasi-continuous electron acceleration in the low corona \citep{kai_storms_1985}.

\subsection{X-ray Imaging: High energy release}

We used multi-viewpoint complementary X-ray observations to pinpoint the location of hot plasma generated by the C3.9-class flare on 6 April. The Spectrometer Telescope for Imaging X-rays onboard Solar Orbiter \citep[SO/STIX;][]{krucker_spectrometertelescope_2020} provided hard X-ray imaging (4--150~keV) sensitive to hot flare plasma ($\gtrsim$10~MK). We reconstructed the flare morphology using SO/STIX forward-fitting routines \citep{volpara_forward_2022}.

Soft X-ray observations from the X-Ray Telescope on Hinode \citep[Hinode/XRT;][]{golub_x-ray_2007} using the Be-med filter provided complementary imaging of thermal emission from $\sim$10~MK flare loops. The combination of SO/STIX and Hinode/XRT observations, obtained from different viewing angles, enabled us to identify the three-dimensional location of the flaring arcade relative to the filament.

\subsection{X-ray Flux: Flare timing and classification}

Flare timing was determined using X-ray flux measurements from the X-ray Sensor (XRS) on the GOES-16 spacecraft \citep{woods_goesr_2024}, which provided the standard GOES classification and temporal profile of the event.

\section{Results}\label{sec:results}

\begin{figure*}
    \centering
    \includegraphics[width=\linewidth]{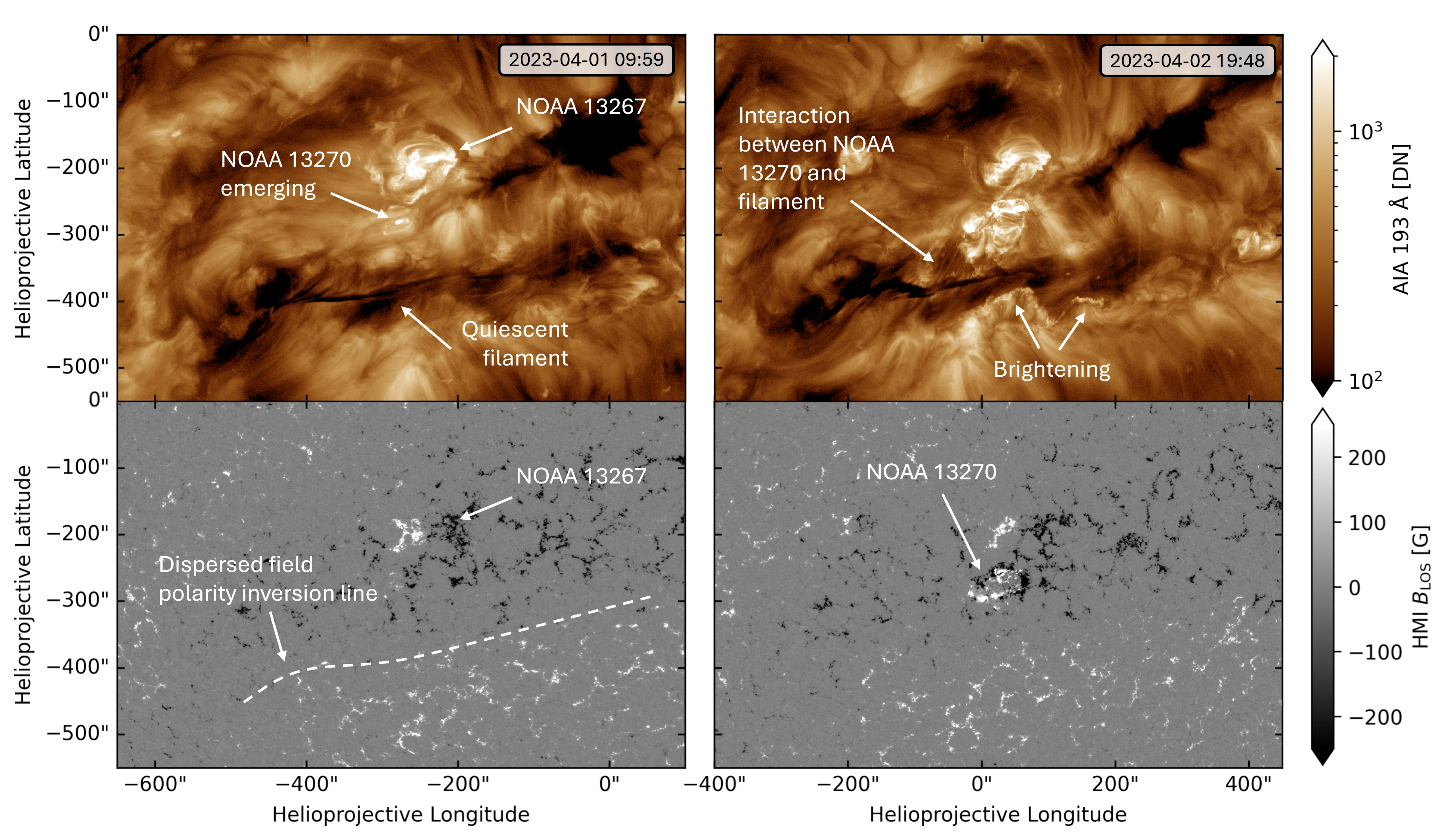}
    \caption{Observations of the emergence of NOAA~13270 on 1~April (left), and the subsequent remote brightenings on 2~April (right). Coronal plasma as seen by SDO/AIA in the 193~\AA{} channel is shown in the top row, and the line-of-sight photospheric magnetic field measured by SDO/HMI is shown in the bottom row.}
    \label{fig:emergence}
\end{figure*}

\subsection{1st and 2nd April: Emergence}

On 1st April 2023, a large quiescent filament extended around 500~arcsec from just south of disk centre to the east, as seen from Earth. The filament channel was formed by a large dispersed positive polarity to the south and a large dispersed negative polarity to the north. Active region NOAA~13267 was on the northern edge of that negative polarity.

At approximately 10:00~UT on 1~April, small-scale EUV activity associated with the emergence of new active region NOAA~13270 was detected. We show this with observations of the coronal plasma in the left panel of Figure~\ref{fig:emergence}. The emergence of NOAA~13270 did not appear to immediately coincide with any notable changes to the surrounding corona, and continued at this low level for approximately 1~day.

Significant magnetic flux emergence forming NOAA~13270 then began at around 10:00~UT on 2~April, and continued for many days. This was not initially accompanied by any noticeable brightening anywhere other than the core of the new active region. By approximately 20:00~UT, continued emergence of photospheric magnetic flux had caused the active region to expand. At the same time, remote brightenings south of the filament were observed. We show these brightenings in the coronal plasma in the right panel of Figure~\ref{fig:emergence}. These brightenings persisted for many days and traced a line mostly parallel with the filament channel on the south side, consistent with the footpoint location of retaining (strapping) unsheared arcades which typically accompany filament channel structures.

At this time, the different active region polarities did not yet form distinct bipoles at the photosphere (see the bottom right panel of Figure~\ref{fig:emergence}). Combined with still being small, this led to the photospheric field generating a complex and low-lying coronal magnetic field topology which dominated the immediately surrounding coronal magnetic field but did not appear to be extensive enough to exert an influence on the surrounding field containing the filament channel.

\subsection{3rd, 4th and 5th April: Entanglement}

\subsubsection{EUV imaging}

\begin{figure*}
    \includegraphics[width=\linewidth]{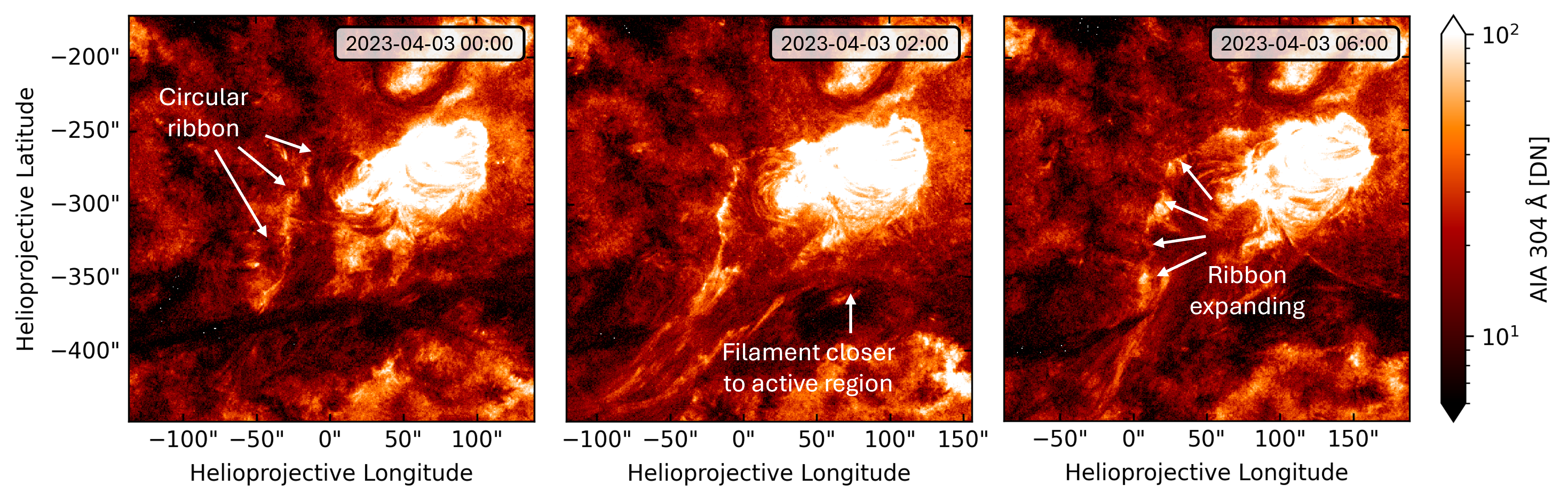}
    \caption{Observations of the expanding circular ribbon around the east side of NOAA~13270, and the movement of the filament material on 3~April, as seen in the SDO/AIA~304~\AA{} channel. This figure is available as an animation.}
    \label{fig:3rdnullevolution}
\end{figure*}

As the active region emergence continued, a partial circular ribbon formed in the chromosphere to the east of NOAA~13270. We show this at 00:00~UT on 3~April in the left panel of Figure~\ref{fig:3rdnullevolution}. The emerging active region is bipolar, where the negative polarity is leading to the west and the positive polarity is trailing to the east. As the active region emerged into a large area of dispersed negative magnetic flux, its eastern trailing positive polarity is, therefore, parasitic in nature.

The circular chromospheric ribbon to the east of the new active region is consistent with the formation of a null point and separatrix dome above the trailing positive polarity, as it is connected to the surrounding negative magnetic flux. Slow steady reconnection at the null point would provide energy to the plasma which would be seen in the ribbon \citep{masson_nature_2009}. As the positive polarity grew with the emergence of more positive photospheric magnetic flux, the boundary of the circular ribbon also grew. This is consistent with the growth of a seperatrix dome encompassing more balancing negative ambient magnetic field.

The middle panel of Figure \ref{fig:3rdnullevolution} shows a strong filament activation taking place shortly after the appearance of the circular ribbon. The filament activation proceeded down the eastern side of the filament originating in the active region. At the same time, dark, cooler filament material is seen displaced closer to the active region. This change of the filament position is not accompanied by any flare or sign of strong reconnection. However, it appears from these observations that active region plasma enters the filament channel at this point of interaction.

Following this, flux emergence continued and the circular ribbon expanded further. At the same time, the filament was disrupted and a complete entanglement of the active region and the filament began. We show this in the right panel of Figure~\ref{fig:3rdnullevolution}. This is consistent with a separatrix fan boundary expanding, meeting and then interacting with the filament channel. At this time, the active region does not cause disruption of the surrounding field sufficient to eject or cancel the filament channel. However, the filament channel does appear to be displaced northward towards the active region.

\subsubsection{NLFFF extrapolations}

\begin{figure}
    \centering
    \includegraphics[width=\linewidth]{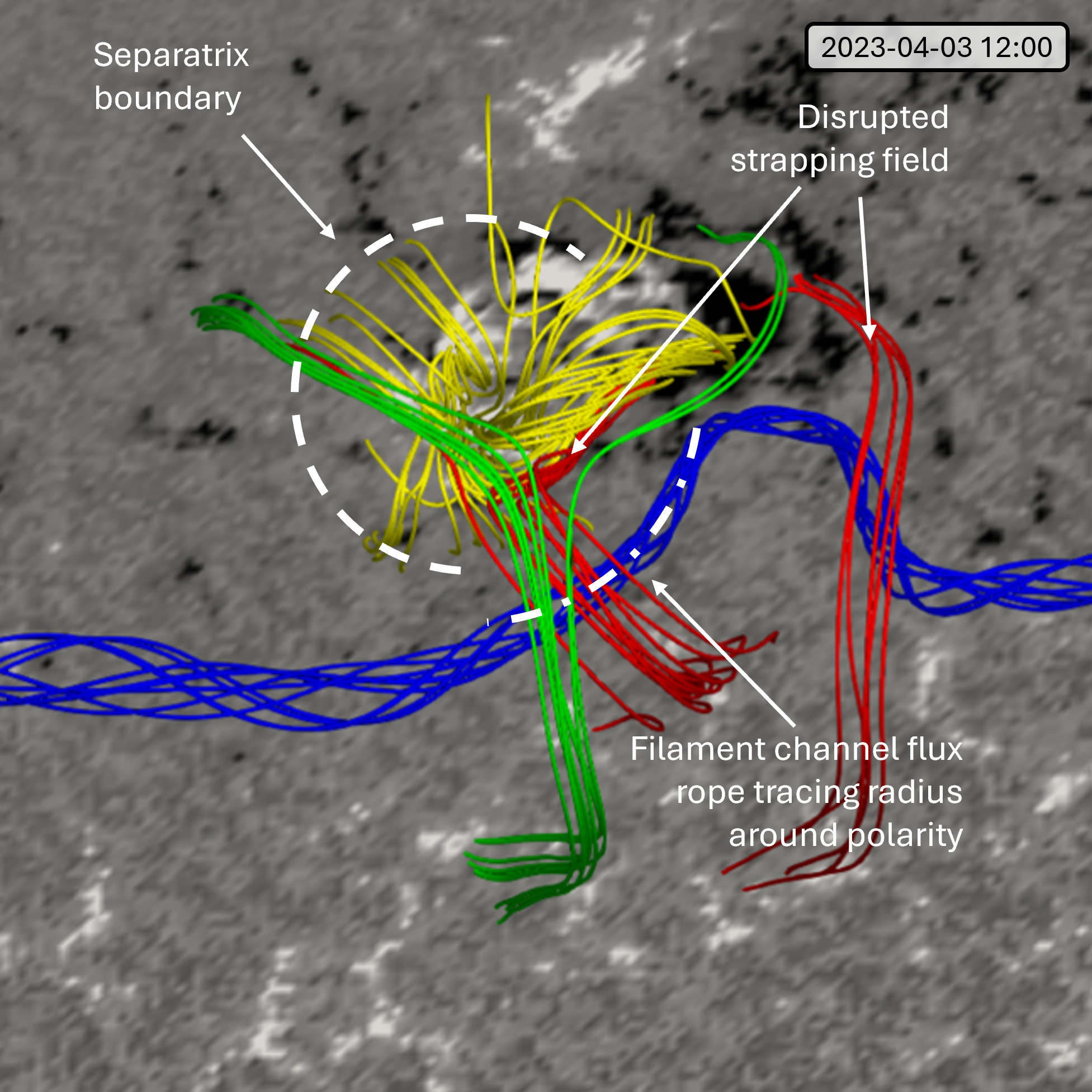}
    \caption{NLFFF extrapolation of the photospheric vector magnetic field measured by SDO/HMI on 3~April. The flux rope at the core of the filament channel is shown in blue, the overlying field in red, the null fan in yellow and the higher strapping field in green. The view point is orientated to match that of Earth.}
    \label{fig:nlfff3rd}
\end{figure}

We perform an NLFFF extrapolation on 3~April to analyse the coronal magnetic topology more closely at this particularly critical point in time. Our extrapolation can be seen in Figure~\ref{fig:nlfff3rd}. Our modelling shows a magnetic fan structure (yellow) around the positive polarity of NOAA~13270 corresponding with the radius traced by the circular ribbons seen in Figure~\ref{fig:3rdnullevolution}. Additionally, the twisted filament channel field lines (blue) also appear to trace a radius around this emerged positive polarity for some distance and towards the active region polarity inversion line, before returning to the field's primary polarity inversion line. The retaining (strapping) high-altitude arcades (green) appear to have separated around the separatrix boundary and the lower-level arcades (red) show shear to match the curvature of the filament channel.

\begin{figure}
  \centering
  \includegraphics[width=\columnwidth]{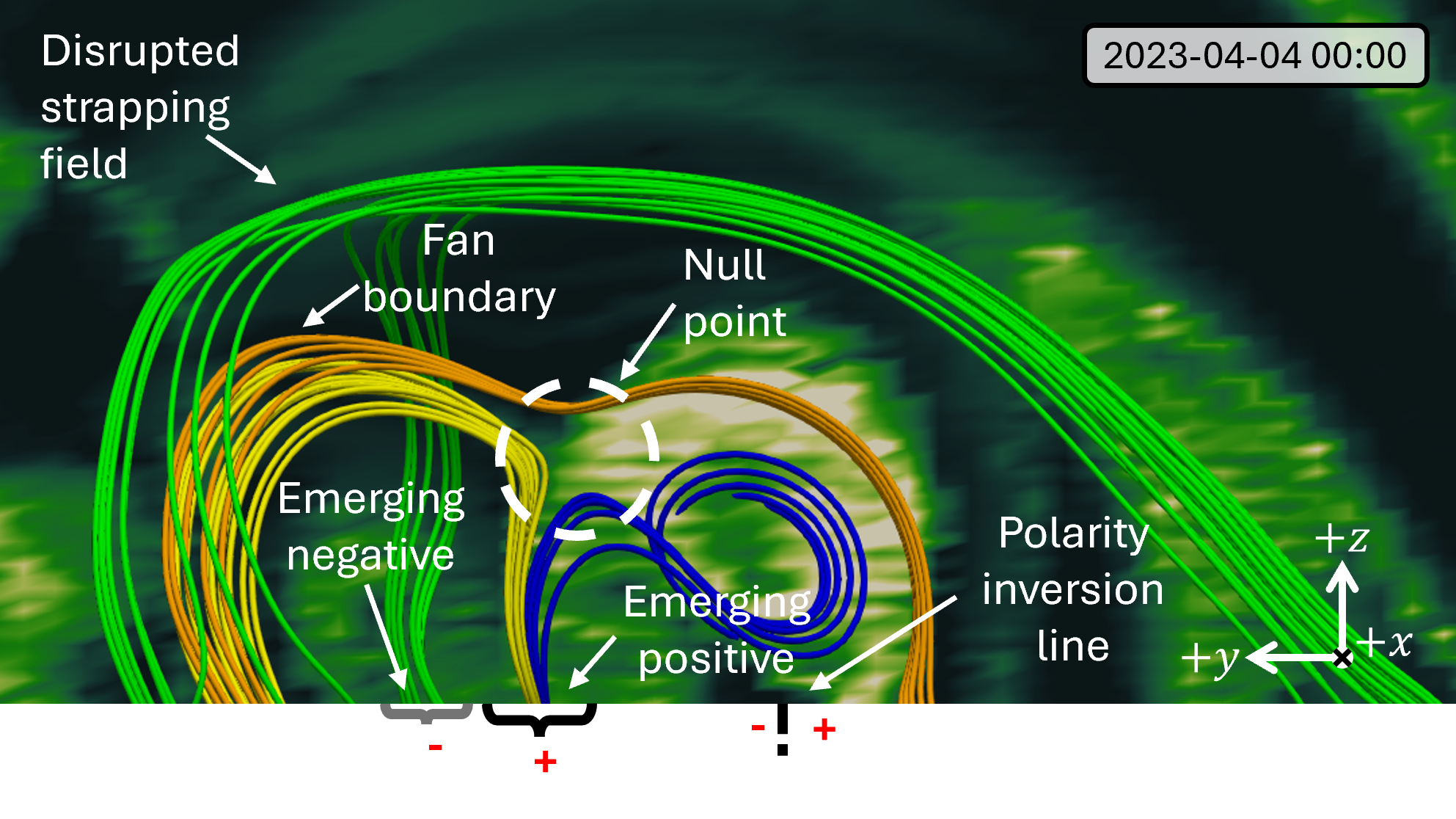}
  \caption{NLFFF extrapolation of the photospheric vector magnetic field measured by SDO/HMI on 4~April. The high altitude strapping field is shown in green, lower-level overlying field in orange, field originating at the new positive polarity and, via a null point forming part of a fan topology in yellow, and field originating in the new positive polarity connected via the null point to the filament channel in blue. A Qmap showing $\log{}Q$ in the y-z plane at the null point is shown, with the colours placed behind the field lines for visibility, and where brighter regions are those of high $Q$ values.}
  \label{fig:null4th}
\end{figure}

Continuing with the analysis of the observations, several hours after this reconfiguration we see remote brightening east of the active region and continued filament activation as active region plasma enters and propagates eastward along the filament channel. In order to investigate possible topological consequences of this process, we performed an NLFFF extrapolation of the region at this time on 4 April and found that the null point was now magnetically connected to the filament channel. We show our extrapolation in Figure~\ref{fig:null4th}. Additionally, the high-altitude filament-retaining (strapping) field (green) continues to be separated above the null fan boundary to the north of the active region, and the lower-level retaining (strapping) field (orange), connected to the photosphere at the south, appears connected to the null point and to the north of the active region.

\subsubsection{Spectroscopic and radio data}

\begin{figure*}
    \centering
    \includegraphics[width=\textwidth]{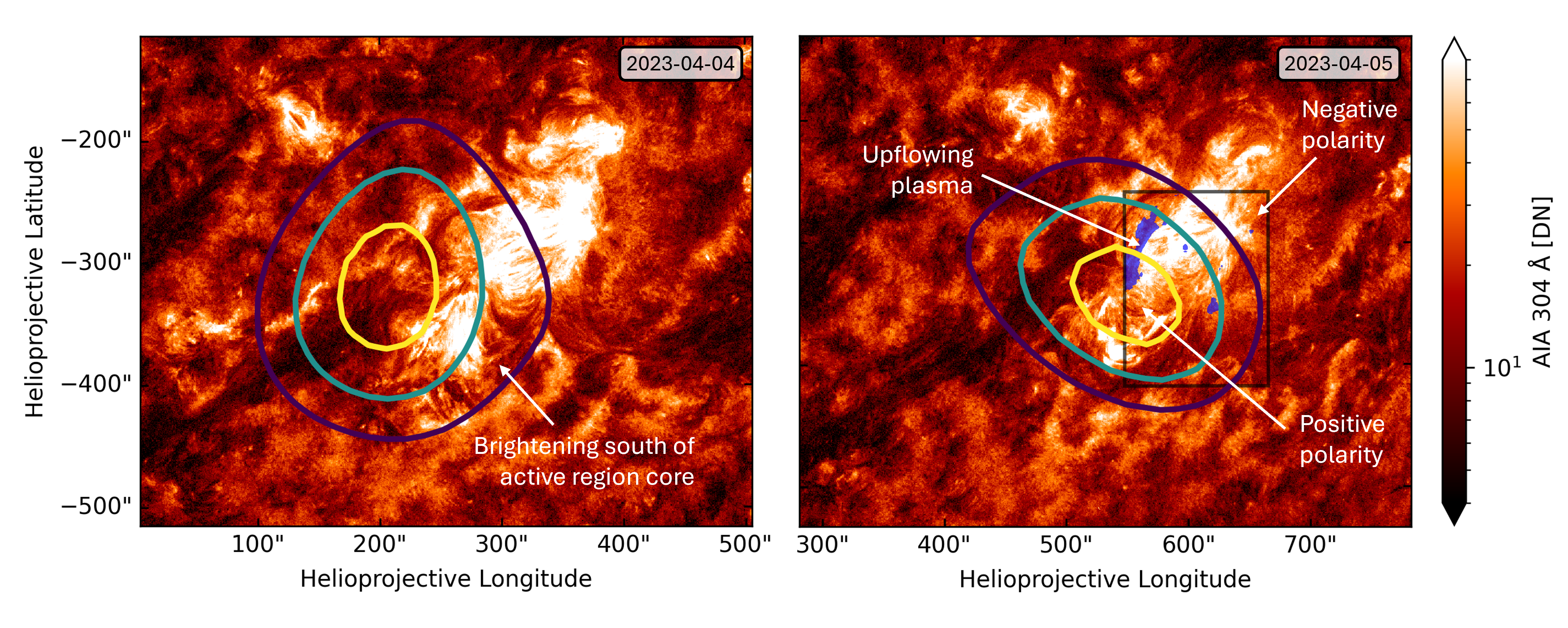}
    \caption{EUV imaging observations of NOAA~13270 using the SDO/AIA 304~\AA{} channel, taken at 07:30~UT on 4~April (left) and 22:05~UT on 5~April (right). Upflowing plasma between 10 and 35~km~s$^{-1}$ observed by Hinode/EIS in \ion{Fe}{12}~195.119~\AA{} is also overlaid in blue (right) at the time of the AIA observation, where the field of view of Hinode/EIS is outlined with a translucent black box. Contours showing 432~MHz radio emission between 75 and 95\% of the peak strength are overlaid. These were taken an 10:01~UT on 4~April (left) and 15:07~UT on 5~April (right), the closest available steady radio data, and corrected for solar rotation.}
    \label{fig:fig_aia_304_04_05_v2_annotated}
\end{figure*}

On 4~April the cool plasma of the filament channel became difficult to trace at the active region location in EUV imaging, and the path of the filament past the active region became unclear. The left panel of Figure~\ref{fig:fig_aia_304_04_05_v2_annotated} shows that a strong source of 432~MHz radio emission on 4~April, the strongest on the whole disk at that time, is concentrated at the same location where the coronal null point was present in our NLFFF extrapolations, above the positive polarity of the active region. The radio signal is indicative of a type~I noise storm that is driven by energetic electrons being quasi-continuously accelerated in the low corona, in areas of plasma density around $10^{9.5}$~cm$^{-3}$ assuming fundamental emission, where the observed radio frequency corresponds to the local plasma frequency \citep[e.g.,][]{dulk_radio_1985}. We do not observe the radio type~I emission at the lower frequencies around 150~MHz, indicating that the accelerated particles do not have access to larger magnetic loops than around 0.4~solar~radii in altitude, based on the plasma density scale height in the corona \citep[e.g.,][]{mann_heliospheric_1999}. The radio emission began on 3~April, but became consistently strong on 4~April. Some low-lying bright loops also appear to connect south from the active region as it further expands. It is at that time that NOAA~13270 had grown to additionally become entangled with NOAA~13267.

Late on 5~April, Hinode/EIS observed persistent plasma upflows of 10--35~km~s$^{-1}$ in the plasma at the coronal \ion{Fe}{12}~195.119~\AA{} (\(\log{T_{\text{max}}=6.2}\)) and hot \ion{Fe}{16}~262.984~\AA{} (\(\log{T_{\text{max}}=6.4}\)) emission lines. The location of the upflowing plasma was identical in both cases, and is shown for \ion{Fe}{12}~195.119~\AA{} in the right panel of Figure \ref{fig:fig_aia_304_04_05_v2_annotated} on 5~April. While slightly affected by the Hinode/EIS field of view, these measurements were sufficient to determine that the upflowing plasma coincided with the strong continuous 432~MHz radio emission and the location of the extrapolated null point. Hinode/EIS continued to observe upflowing plasma of similar velocity from this same location until 7~April when the active region was too close to the limb to be followed further.

5~April was the start of a more severe displacement of the eastern side filament away from its original location, seen in the right panel of Figure~\ref{fig:fig_aia_304_04_05_v2_annotated}. It is unclear from observations at this point in time, given the brightness of the active region plasma, whether the filament channel maintains an uninterrupted coherent structure with a single axis past or through the active region core, or whether there is some disconnection between the east and west sides of the filament, with the west side remaining more undisturbed than the east. We speculate further over such possibilities in Section \ref{sec:discussion}.

\subsection{6th and 7th April: Jets and Flare}

\begin{figure}
  \centering
  \includegraphics[width=\columnwidth]{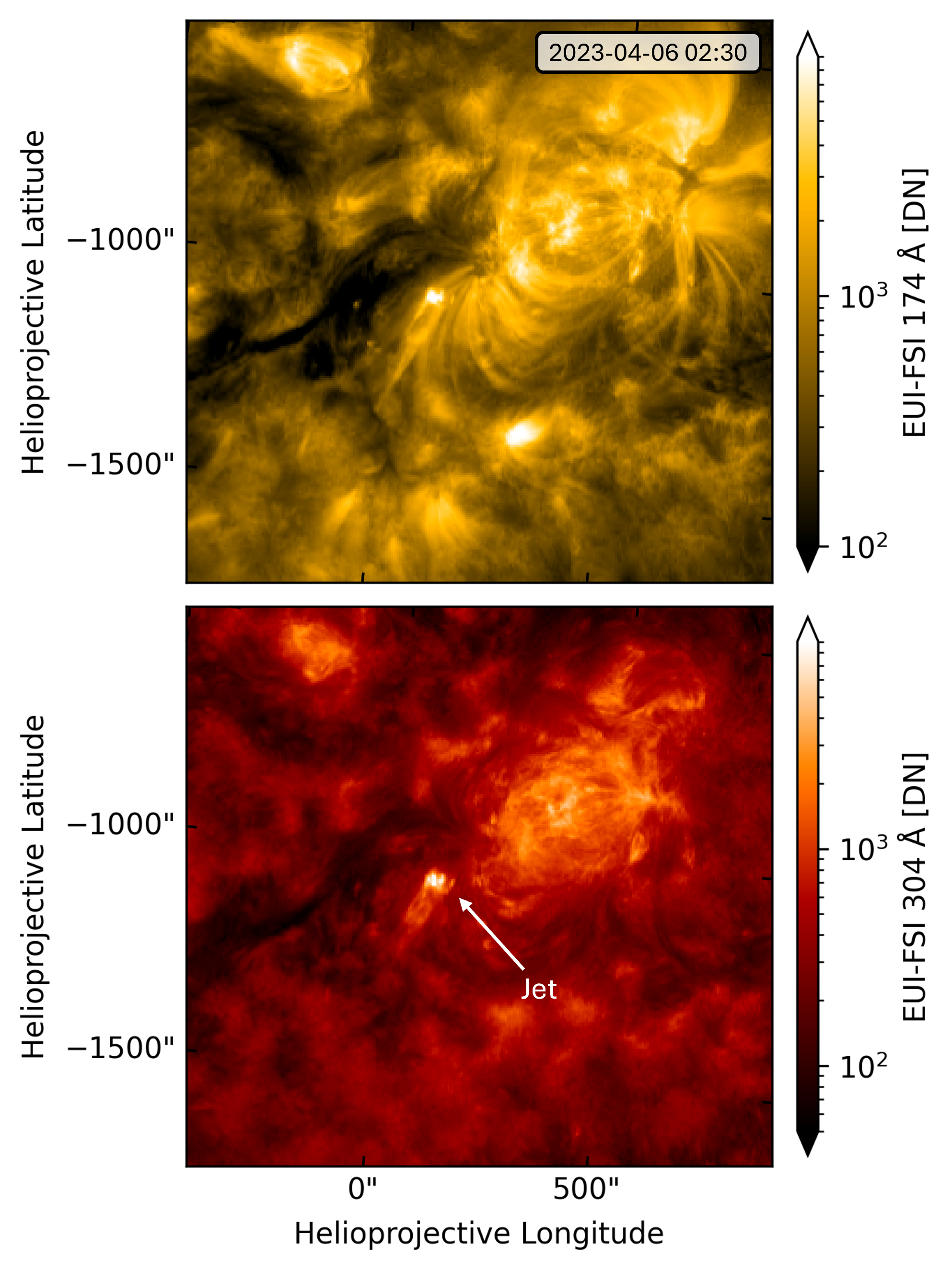}
  \caption{Observations of a jet on the east side of NOAA~13270 as seen by SO/EUI-FSI.}
  \label{fig:jet06}
\end{figure} 

Early on 6~April, the relative position of the filament material to the east and west of the active region was largely unchanged. Using EUV imaging, it remains indiscernible whether a continual filament channel passes through or past the active region, or whether what we see are simply the two ends of new, separate filament channels. The filament material on the eastern side of the active region remained elevated and more northern, while the west filament material remained radially above the ambient field's polarity inversion line. Around this time, numerous jets are seen on the eastern side of the active region, an example of which we show at 02:30~UT on 6~April in the bottom panel of Figure \ref{fig:jet06}. The jet appears to originate slightly south of the filament material to the east of the active region.

\begin{figure*}
    \centering
    \includegraphics[width=\linewidth]{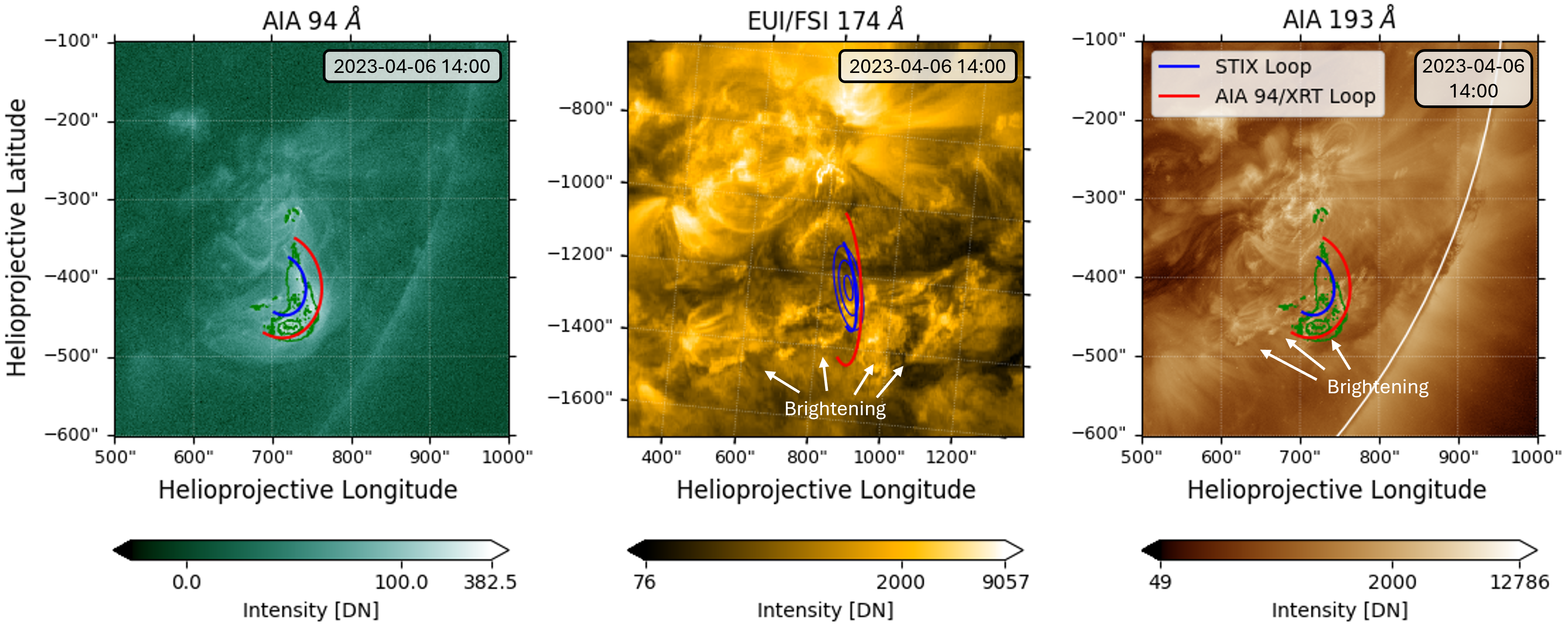}
    \caption{EUV imaging observations of the C3.9 flare on 6~April seen by the SDO/AIA 94~\AA{} (left) and 193~\AA{} (right) channels, and around the Sun from Earth by SO/EUI-FSI~174~\AA{} (centre). We overlay intensity contours at 30\%, 60\% and 90\% of the peak value from Hinode/XRT in green, and those same contours for SO/STIX in blue. We overlay the flaring loop identified by SDO/AIA~94~\AA{} and Hinode/XRT in red, and the flaring loop identified by SO/STIX in blue.}
    \label{fig:flare}
\end{figure*}

At approximately 14:00~UT on 6~April, a GOES C3.9-class flare occurred. It lasted until 14:29~UT and caused associated EUV brightenings in the surrounding plasma that persisted for several hours. To determine the 3D location of the energy release relative to the filament, we combined EUV imaging with X-ray observations from two view points at Earth and from Solar Orbiter further around the Sun (see Figure~\ref{fig:sc_locs}).

We show the flare topology in Figure~\ref{fig:flare} by overlaying the location of the hot X-ray emitting plasma onto images of EUV emission. The left and right panels show the view from Earth. Here, the soft X-ray emission observed by Hinode/XRT (green contours) identifies the hot thermal plasma. This emission is co-spatial with hot loop structures seen in the SDO/AIA~94~\AA{} channel (\(\log{T\sim{}6.8}\)), and so we annotate all the panels of this figure with this AIA~94~\AA{}/XRT loop in red. The central panel shows the view from Solar Orbiter using SO/EUI-FSI~174~\AA{}. The blue contours in this panel represent the location of the peak X-ray intensity reconstructed from SO/STIX data. We plot the 3D loop geometry derived from the SO/STIX forward-fitting analysis on all panels with a blue line, which aligns well with the AIA~94\AA{}/XRT loop.

The combined viewing angles strongly suggest that the flaring structure is an arcade overlying the filament channel, rooted in the dispersed positive field south of the filament and in negative polarity of the active region to the north. This is consistent with energy deposition at the arcade footpoints seen in the brightening EUV ribbons parallel to, and south of, the filament which appeared at the same time as the flare, shown in the centre and right panels.

The filament remained stable in position despite the flare.

\section{Discussion}\label{sec:discussion}

\subsection{Topology of emerging active region}

The interaction between the emerging bipolar active region NOAA~13270 and the pre-existing quiescent filament channel shows a complex evolution of topology. Our observations and extrapolations consistently show that a fan-spine separatrix layer formed around the positive polarity of NOAA~13270, which acted as a parasitic polarity embedded in the surrounding ambient negative field. In the extrapolation, the fan separatrix also appeared to exhibit an asymmetry in the density of its magnetic footpoints at the photosphere, which can be expected to yield a correspondent asymmetry in the energy density deposition from the reconnection outflows along the fan. This is consistent with the observed appearance of partial circular ribbons in the chromosphere.

As magnetic flux continued to emerge, we observed signatures of the expansion of the separatrix fan, causing a lateral expansion of the partial circular ribbons to progressively encompass more of the surrounding dispersed negative magnetic flux.

Reconnection at the null point, as described by \cite{pontin_generalised_2011}, likely provided a source of persistent energy release into the coronal plasma throughout the period studied here, which we observed as chromospheric brightening at the fan's footprint and, once connected to the filament channel, as a continual source of filament activation. Importantly, the null point did not collapse into a current sheet according to our NLFFF extrapolations, as has been seen in the past \citep{pontin_three-dimensional_2011}.

\subsection{Filament activation and magnetic reconfiguration}

The growing of the separatrix fan then likely directly interacted with the adjacent filament channel once expanded sufficiently, inducing the significant filament activation seen by hot plasma propagating through the filament from the active region. This represented the first of several perturbations caused by the emerging active region without causing the filament eruption or complete reconfiguration.

Our coronal field extrapolations found the filament's magnetic axis curved around the active region's positive polarity, deviating from its initial alignment along the ambient field's polarity inversion line. As we discuss in more details in Section~\ref{sec:cartoon_model}, this bending may have arisen because the parasitic polarity displaced the strapping field's anchoring region on one side of the filament channel, forcing reconfiguration and reducing magnetic pressure on the northern side of the filament. This reduction was balanced by the separatrix layer's magnetic pressure at the positive polarity, while at the negative polarity, the strapping field became sheared, drawing the filament closer to the active region. The accompanying remote brightenings, south of the active region, then occurred in the same polarity as that of the parasitic polarity, as in the case seen by \cite{wang_circular_2012}. This partial tracing of a radius around the positive parasitic polarity is similar to the circular filament observed by \cite{dacie_sequential_2018}, in that twisted filament channel field lines can curve along their axial direction around a parasitic polarity.

The same magnetic field extrapolation found overlying strapping field separated on the northern side of the filament, logically as the southern-rooted field lost its northern counterpart due to the emerged negative polarity. Such an opening of magnetic flux can facilitate filament eruption \citep[e.g.,][]{wang_filament_1999}. However, the filament did not erupt meaning no large-scale eruptive mechanism, such as breakout \citep{antiochos_magnetic_1998} or tether-cutting \citep{svestka_triggering_1992}, was successfully triggered.

\subsection{Radio emission and persistent plasma upflows}

Following entanglement with the filament channel, the active region null point appeared to persist, coinciding with strong and persistent radio emission at 432~MHz. Similar radio signatures observed at null points by \cite{duan_homologous_2022} were attributed, as here, to slow, steady reconnection at such sites. This reconnection would provide continual particle acceleration that drives radio emission such as we see here, along with heating that would drive ongoing filament activation and chromospheric brightening at the separatrix fan's footprint, which we observed.

Persistent plasma upflows (10--35~km~s$^{-1}$) were observed in coronal (\ion{Fe}{12}) and hotter (\ion{Fe}{16}) emission lines at the location of the radio emission and near the likely null point. While such upflows are routinely associated with active regions \citep[e.g.,][]{brooks_formation_2021, tian_upflows_2021}, in this case it is noteworthy that strong upflows persist despite significant interaction with a large-scale filament channel, which could reasonably be expected to influence or disrupt typical active-region dynamics. The persistence of these strong upflows indicates that active regions inherently produce such flows even when embedded within complex, larger-scale coronal topologies.

While the plasma upflows showed structure broadly co-spatial with the separatrix layer we suggest is present, Hinode/EIS's limited spatial resolution makes it challenging to pinpoint the precise mechanism driving them conclusively. Future plasma spectrometers like SOLAR-C/EUVST will enable observations of such dynamics at unprecedented spatial resolution, significantly improving our understanding of these dynamics \citep{mckevitt_pre-flare_2025,shimizu_solar-c_euvst_2019}.

\subsection{Jets as mini-filament eruptions}

We observed numerous jets adjacent to the filament-active region complex following their interaction. \citet{sterling_small-scale_2015}, supported by \citet{wyper_universal_2017}, proposed that such jets represent small-scale eruptions of sheared magnetic fields (\lq{}mini-filaments\rq{}) meaning that in our case, such sheared field is continually ejected from the region. These frequent, small eruptions may effectively relieve the accumulated magnetic shear and currents from the coronal field, which would contribute to the stability of the larger filament structure which we see here.

\subsection{Flare dynamics and filament stability}

A C-class flare on 6~April represented the peak energy release during our observations but notably was not associated with filament eruption. X-ray and EUV imaging (SDO/AIA, SO/STIX, Hinode/XRT) localised the flare within the strapping field overlying the filament rather than within highly sheared fields beneath it, as would be the case for an eruption \citep{svestka_triggering_1992, antiochos_magnetic_1998}.

\subsection{Filament integrity and potential fragmentation}

As the region moved to the edge of the limb beyond 6~April, despite prolonged magnetic perturbations, it remains unclear whether the filament maintained continuity across the active region or fragmented into two intermediate filaments anchored near the active region. EUV observations indicated elevation and northward displacement of the eastern filament segment, suggesting possible partial disconnection or weakening of filament integrity. Such fragmentation reflects nuanced filament responses to strong but localised perturbations, complicating eruption predictions.

The question of filament continuity remains open due to the limitations of magnetic field extrapolation methods at significant distances from the disk centre. This highlights the need for high-resolution, multi-viewpoint magnetic observations (e.g., via SO/PHI-FDT; \citealt{valori_stereoscopic_2023}) to resolve the detailed quiet Sun photospheric magnetic field underpinning such scenarios.

\subsection{Comparison to previous studies and theoretical implications}

\cite{feynman_initiation_1995} described emergences into dispersed single-polarity regions, such as we consider here, as neither favourable nor unfavourable for reconnection, events equally likely to erupt or remain stable and relatively rare observationally.

The interaction studied here aligns partially with previous flux emergence reports \citep[e.g.,][]{dacie_sequential_2018, torok_solar_2024, wang_filament_1999} but notably diverges by not producing a large-scale eruption. The work by \cite{dacie_sequential_2018} on an eruptive case considers emerging elongated bipoles with a polarity inversion line parallel to that of the ambient field. This contrasts with our non-eruptive case in which the active region polarity inversion line is perpendicular to the ambient field's.

A statistical study of those emergences into the dispersed single-polarity regions referred to by \citet{feynman_initiation_1995}, considered alongside the orientation of the polarity inversion line of the emerging field with respect to that of the ambient field, would be interesting to see if any correlation is present.

Our observations provide an interesting perspective on the two-step reconnection scenario proposed by \cite{wang_flare-associated_1993} wherein continuous, relatively slow reconnection, observed as flux cancellation at the photosphere, builds magnetic complexity into large magnetic structures in the corona before energy release takes place in the corona. In our case, continuous reconnection at the null point releases energy from the magnetic field without triggering an eruption, and jets potentially release shear from the field without an eruption of the large-scale filament.

\subsection{Cartoon model}\label{sec:cartoon_model}

\begin{figure}
  \centering
  \includegraphics[width=.9\columnwidth]{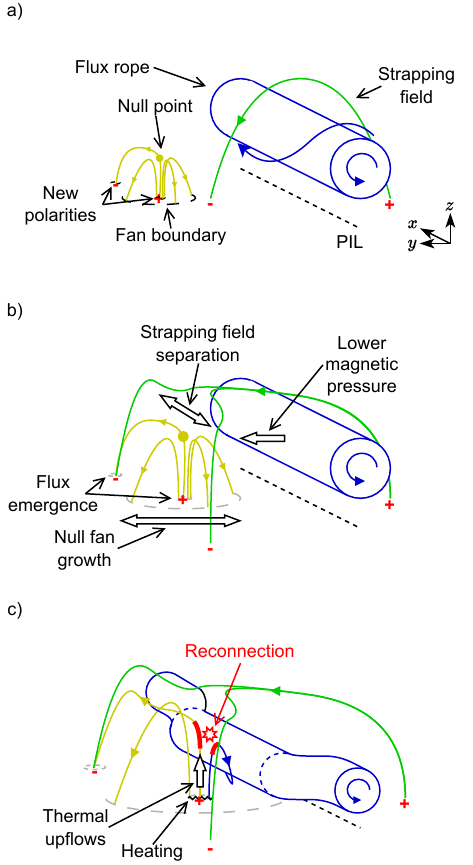}
  \caption{A proposed cartoon model for the interaction between the active region and filament channel. The filament channel (flux rope; blue) initially resides above the ambient field's polarity inversion line (PIL) next to a null fan (yellow) in (a), the strapping field (green) is separated (b), and the channel is then seen connected to the null point reconnection site (c). The colours correspond to those used in Figures \ref{fig:nlfff3rd} and \ref{fig:null4th}.}
  \label{fig:cartoon}
\end{figure}

To illustrate our interpretation of the observed interactions, we present a cartoon model (Figure~\ref{fig:cartoon}) corresponding to the observed coronal plasma and the magnetic topologies we model in Figures~\ref{fig:nlfff3rd} and \ref{fig:null4th}

In panel (a), we show the initial magnetic configuration prior to significant interaction. A flux rope associated with the filament channel lies along a dispersed field polarity inversion line, stabilised by strapping fields from the surrounding ambient field. The emergence of new magnetic flux from NOAA~13270 introduces a quasi-parasitic polarity, forming a fan-spine topology characterised by a dome-shaped separatrix surface (the fan) and an associated coronal null point. This newly emerged configuration, embedded within the existing dispersed negative polarity region, creates a coronal null point offering a favourable site for reconnection.

Panel (b) illustrates the expansion of the null fan as continued flux emergence injects positive polarity flux into the corona. This rapid expansion disrupts and separates portions of the strapping field originally stabilising the filament channel. Importantly, the remaining connected strapping field lines still anchor into the dispersed negative field region, thus drawing the filament channel field lines beneath the expanding null fan surface, towards the region of lower magnetic pressure generated by the emergence.

Finally, in panel (c), we illustrate the resulting persistent magnetic reconnection occurring near the null point, causing upflowing plasma and continuous particle acceleration. The connection between this reconnection site and the filament channel is responsible for the continual activation of the filament.

\section{Conclusions}\label{sec:conclusions}

In this work we analyse why a quiescent filament remained stable despite major magnetic disturbances from the emergence of NOAA~13270.

The first signatures of the emergence of active region NOAA 13270 on 1~April 2023 produced localised EUV activity and no disturbance to the large nearby quiescent filament. By 2~April following the emergence of more photospheric magnetic flux, we observed remote EUV brightenings on the opposite side of the filament indicating that the emerging flux had begun interacting with the unsheared arcades retaining the filament. As the emergence continued, we identified the appearance and expansion of a partial circular ribbon near NOAA~13270. We find its behaviour is consistent with the formation of a separatrix structure around the emerging positive polarity, which acted as a parasitic polarity within the surrounding negative magnetic flux.

By 3--4 April, the filament and emerging flux became fully entangled. We performed NLFFF magnetic field extrapolations and found a seperatrix surface and associated null point above the emerging positive polarity, consistent with the observed circular ribbon. Our extrapolations also found magnetic connectivity between this inferred null point and the filament channel itself, consistent with the signatures we observed of the apparent injection of hotter plasma into the filament channel from the active region. We also detected a strong, spatially stable 432~MHz radio source above the emerging positive polarity, indicative of particle acceleration caused by ongoing magnetic reconnection at a null point.

Later, on 5~April, Hinode/EIS measurements of bulk plasma motion found coronal and hot-plasma upflows of 10--35~km~s$^{-1}$ co-spatial with the radio source and the extrapolated null point location. These upflows are consistent with long-duration, gradual energy release. At this time the filament was still in place and largely stable with the active region embedded in it despite the emergence of flux at the photosphere and growth of NOAA~13270, ongoing reconnection at the null, and coronal activity. On 6~April, a sequence of jets occurred near the filament--active region interface. Current literature suggests such jets are the ejection of \lq{}mini-filaments\rq{}. This would help explain the non-eruption of our filament in this case in that the jets may have contributed to relieving magnetic stress from the system.

Our multi-viewpoint analysis of a confined C3.9 flare on 6~April finds it to be located in the overlying strapping loops not beneath the filament, likely contributing to the stability of the filament during the event.

At every stage, the system appears to have evolved through slow reconnection, topological adjustment, and shear-relieving small-scale ejection, rather than the explosive mechanisms required to destabilise and erupt the filament. In comparison with previous eruptive studies, the emerging flux here was perpendicularly orientated relative to the filament channel, something we suggest is an important parameter in determining the stability of the filament.

\begin{acknowledgments}

We are thankful to the referee for the comments and suggestions that helped to improve the manuscript. J.M. was supported by STFC PhD Studentship number ST/X508858/1. S.L.Y. is grateful to the Science Technology and Facilities Council for the award of an Ernest Rutherford Fellowship (ST/X003787/1). H.R. was supported by UKSA grant ST/X002012/1 and H.R. and A.W.J. by STFC grant ST/W001004/1. S.M. and D.B. acknowledge support from UKSA grant No. UKRI920. S.M. was also supported by ESA Contract No. 4000141160/23/NL/IB. R.M.D. was supported by NASA contract NNG10EK25C. L.Z. acknowledges support from NASA grant 80NSSC22K1015, NSF SHINE grant 2229138, and NSF Early Career grant 223743. The work of D.H.B. was performed under contract to the Naval Research Laboratory and was funded by the NASA Hinode program. The authors acknowledge the use of data from the Solar Dynamics Observatory (SDO). Courtesy of NASA/SDO and the AIA, EVE, and HMI science teams. Solar Orbiter is a mission of international cooperation between ESA and NASA, operated by ESA. Hinode is a Japanese mission developed and launched by ISAS/JAXA, with NAOJ as domestic partner and NASA and STFC (UK) as international partners. It is operated by these agencies in co-operation with ESA and NSC (Norway). NLFFF extrapolations and data analysis were performed on the Vienna Scientific Cluster (VSC; https://vsc.ac.at/). We thank the Solar group at MSSL for their discussions which helped progress this work.

\end{acknowledgments}

\appendix

\section{Magnetic field extrapolations}

\begin{figure}
    \centering
    \includegraphics[width=\linewidth]{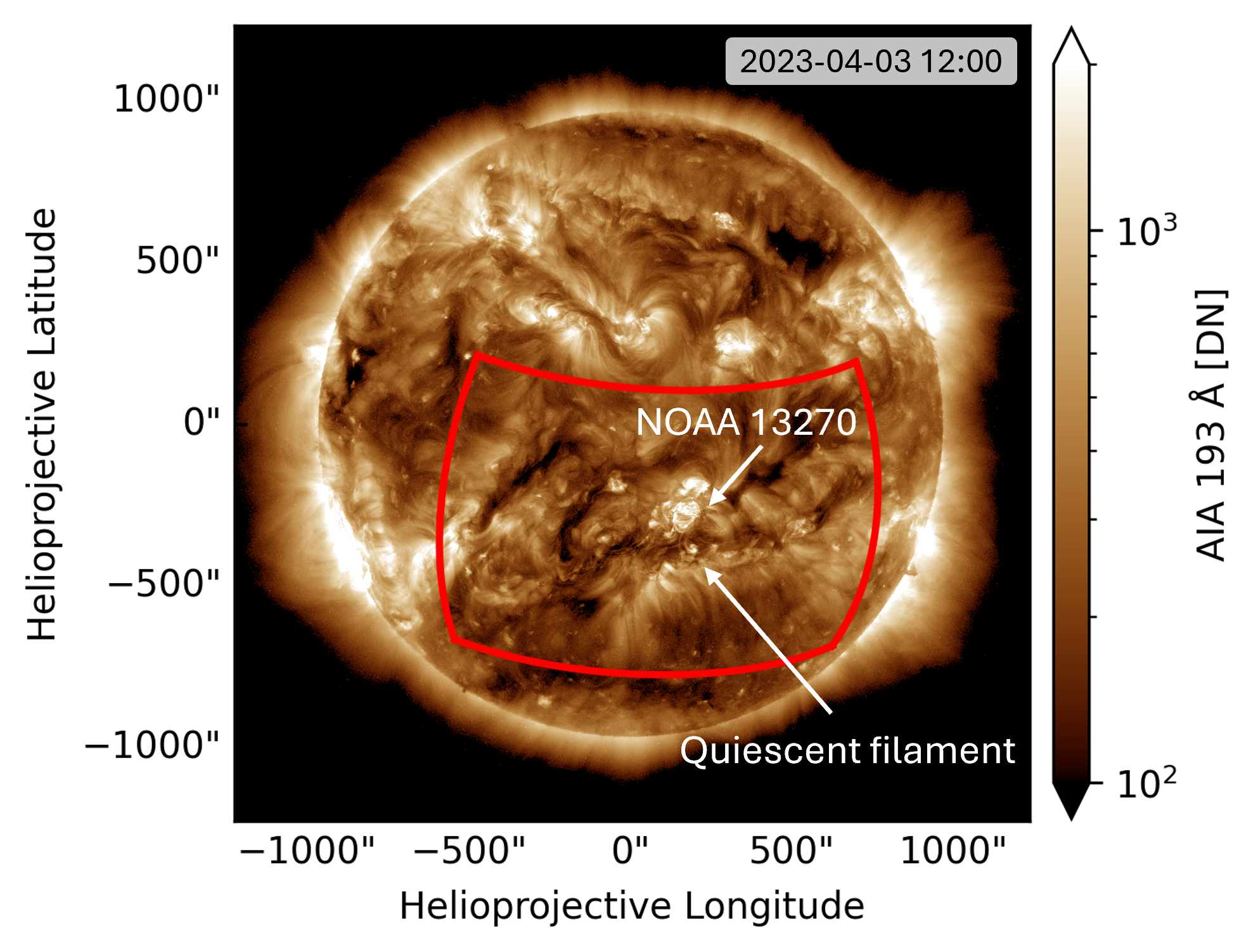}
    \caption{The region of the disk for which NLFFF modelling was performed outlined with a red box. The disk is shown as seen by the SDO/AIA 193~\AA{} channel.}
    \label{fig:extrap_region}
\end{figure}

We used the flux-conserving spherical polygon intersection reprojection method of \cite{the_astropy_collaboration_astropy_2022} to generate a grid of photospheric fluxes in Cartesian cylindrical equal area (CEA) coordinates at the bottom boundary of the extrapolation from SDO/HMI data. The size of the domain is large in order to capture the dispersed field associated with the quiescent filament (Figure~\ref{fig:extrap_region}), and the reprojection to a Cartesian CEA system is an approximation of a spherical coordinate domain. As we were primarily concerned with the central region away from any slightly distorted edges, we find this approximation sufficient. We then performed a potential field extrapolation using the Green's function of \cite{sakurai_greens_1982} to provide boundary conditions for the other (side and top) boundaries.

In a volume within these boundaries, we consider a force-free field \citep[e.g.,][]{wiegelmann_solar_2021}, where $\mathbf{J} \times \mathbf{B} = \mathbf{0}$ and $\nabla \cdot \mathbf{B} = 0$, $\mathbf{J}$ being the electric current and $\mathbf{B}$ being the magnetic field. To generate a field which satisfies these equations, we use a physics-informed neural network (PINN) method \citep{raissi_physics-informed_2019}, adapted from \cite{jarolim_probing_2023}. See \cite{baty_modelling_2023} for a discussion of the intricacies of extrapolating coronal magnetic fields with PINN methods. When performing a direct extrapolation of \(\mathbf{B}\) using this method, as in \cite{mckevitt_link_2024}, we found difficulties maintaining \(\nabla\cdot{}\mathbf{B}=0\). We therefore implemented a vector potential approach, similar to the approach of \cite{jarolim_magnetic_2024} but with some adjustments. In such a method, $\nabla \cdot \mathbf{B} = 0$ is inherently satisfied by representing the magnetic field as the curl of a vector potential field, \(\mathbf{A}\):

\begin{equation}
    \mathbf{B} = \nabla \times \mathbf{A}.
    \label{equ:B_from_A}
\end{equation}

In this paper, we generate this \(\mathbf{A}\) field with our solver and minimise a loss function from the subsequently calculated \(\mathbf{B}\) field. Since the divergence of a curl is identically zero, this representation guarantees $\nabla \cdot \mathbf{B} = 0$ by construction.

An unconstrained vector potential $\mathbf{A}$ can be augmented by an arbitrary scalar field $\chi$ such that:

\begin{equation}
    \mathbf{B}=\nabla\times\mathbf{A}=\nabla\times\left(\mathbf{A}+\nabla\chi\right),
\end{equation}

\noindent{}given that $\nabla\times\nabla\chi=0$ \citep[Equation 15,][]{beresnyak_nrl_2023}. To handle this \lq{}gauge freedom\rq{} inherent in the vector potential, we impose the Coulomb gauge condition \(\nabla \cdot \mathbf{A} = 0\), which we enforced through a Coulomb gauge loss term:

\begin{equation}
    \mathcal{L}_{\text{CG}} = \frac{1}{N_{\text{CG}}} \sum_{i=1}^{N_{\text{CG}}} \left( \nabla \cdot \mathbf{A}(\mathbf{r}_{i}) \right)^2,
    \label{equ:coulomb_gauge_loss}
\end{equation}

\noindent{}where \(N_{\text{CG}}\) is the number of collocation points for the gauge condition and \(r=(x, y, z)\). These collocation points are randomly-sampled locations in the spatial domain where the neural network is penalised for deviating from the constraining equations.

Enforcing the Coulomb gauge also allows us to define the force-free loss (see \cite{valori_magnetic_2016} for details) as:

\begin{equation}
    \mathcal{L}_{\text{FF}} = \frac{1}{N_{\text{F}}} \sum_{i=1}^{N_{\text{F}}} \left|\left| \left( -\nabla^{2} \mathbf{A} \left( \mathbf{r}_{i} \right) \right) \times \left( \nabla \times \mathbf{A} \left( \mathbf{r}_{i} \right) \right) \right|\right|^{2},
    \label{equ:force_free_loss}
\end{equation}

\noindent{}where $N_{\text{F}}$ is the number of collocation points for the force-free calculation.

We calculate a boundary loss term $\mathcal{L}_{\text{B}}$, as in \citep{jarolim_probing_2023}, by taking the mean of all squared differences between the input magnetogram and the bottom boundary of the computed field. We then define our total loss function as:

\begin{equation}
    \mathcal{L} = \lambda_{\text{B}} \mathcal{L}_{\text{B}} + \lambda_{\text{FF}} \mathcal{L}_{\text{FF}} + \lambda_{\text{CG}} \mathcal{L}_{\text{CG}},
    \label{equ:total_loss}
\end{equation}

\noindent{}where \(\lambda_{\text{i}}\) are the respective weighting factors.

The training of the neural network involves minimising the total loss function using the Adam optimiser \citep{kingma_adam_2015}. When sampling points in our spatial domain for the forward pass, loss computation, backward pass and network update, we introduce a sampling of collocation points which is gently weighted towards stronger magnetic field locations. This allowed us to better minimise the loss function in the key structures of our large domain.

We evaluate the correctness of our solutions in both their force-freeness and divergence-freeness using the $\theta_J$ and $f_i$ terms of \cite{wheatland_optimization_2000}. Across our entire simulation volume for all our extrapolations, we found \(\theta_J\leq{}23^{\circ}\) and \( \langle|f_i|\rangle\leq{}7.5\times10^{-6}\), lying in an acceptable range \citep[e.g.,][]{derosa_critical_2009}.

\bibliographystyle{aasjournal}
\bibliography{references}

\end{document}